\documentclass[useAMS,usenatbib,a4paper,preprint]{mn2e}
\usepackage{graphicx}
\usepackage{amsmath}
\usepackage{txfonts}
\usepackage{mathrsfs}
\usepackage{color}

\def\aap{A\&A}
\def\apjl{ApJ}

\def\apjs{ApJS}

\newcommand{\be}{\begin{equation}}
\newcommand{\ee}{\end{equation}}
\newcommand{\bea}{\begin{eqnarray}}
\newcommand{\eea}{\end{eqnarray}}

\title[Compact Quasar Cores]{The Maximum Angular-Diameter Distance in Cosmology} 
  \author[Fulvio Melia \& Manoj Yennapureddy]
    {Fulvio Melia$^1$\thanks{John Woodruff Simpson Fellow. Email: fmelia@email.arizona.edu}
    and Manoj K. Yennapureddy$^{2}$\thanks{manojy@email.arizona.edu} \\
     $^1$Department of Physics, The Applied Math Program, and Department of Astronomy,
     The University of Arizona, AZ 85721, USA \\
     $^2$Department of Physics, The University of Arizona, AZ 85721, USA }

\begin{document}

\date{}

\pagerange{\pageref{firstpage}--\pageref{lastpage}} \pubyear{2016}

\maketitle

\label{firstpage}

\begin{abstract}
Unlike other observational signatures in cosmology, the angular-diameter 
distance $d_A(z)$ uniquely reaches a maximum (at $z_{\rm max}$) 
and then shrinks to zero towards the big bang. The location of this turning point
depends sensitively on the model, but has been difficult to measure. In this paper, 
we estimate and use $z_{\rm max}$ inferred from quasar cores: (1) by employing 
a sample of 140 objects yielding a much reduced dispersion due to pre-constrained 
limits on their spectral index and luminosity, (2) by reconstructing $d_A(z)$ using 
Gaussian processes, and (3) comparing the predictions of seven different
cosmologies and showing that the measured value of $z_{\rm max}$ can 
effectively discriminate between them.  We find that $z_{\rm max}=1.70
\pm0.20$---an important new probe of the Universe's geometry. 
The most strongly favoured model is $R_{\rm h}=ct$, followed by {\it Planck} 
$\Lambda$CDM. Several others, including Milne, Einstein-de Sitter and Static tired 
light are strongly rejected. According to these results, the $R_{\rm h}=ct$ universe, 
which predicts $z_{\rm max}=1.718$, has a $\sim 92.8\%$ probability of being the 
correct cosmology. For consistency, we also carry out model selection based on 
$d_A(z)$ itself. This test confirms that $R_{\rm h}=ct$ and {\it Planck} 
$\Lambda$CDM are among the few models that account for angular-size data
better than those that are disfavoured by $z_{\rm max}$. The $d_A(z)$ comparison, 
however, is less discerning than that with $z_{\rm max}$, due to the additional
free parameter, $H_0$. We find that  $H_0=63.4\pm1.2$ km s$^{-1}$ Mpc$^{-1}$ for 
$R_{\rm h}=ct$, and $69.9\pm1.5$  km s$^{-1}$ Mpc$^{-1}$ for $\Lambda$CDM. Both 
are consistent with previously measured values in each model, though they differ from 
each other by over $4\sigma$. In contrast, model selection based on $z_{\rm max}$ is 
independent of $H_0$.

\end{abstract}


\section{Introduction}
Several attempts to measure the angular diameter distance, $d_A(z)$,
 have been made over the past three decades, but so far only with limited
success due to a lack of evident `standard rulers' and limitations from
possible size evolution with redshift. Recently, however, our understanding 
of compact quasar cores has improved to the point where one may now
use the central, opaque regions in a luminosity-constrained sample 
as reliable measuring rods. These show negligible redshift evolution 
in the range $0\lesssim z\lesssim 3$, allowing us to examine the geometry 
of the Universe over an even larger fraction of its age than is possible 
with Type Ia SNe. 

Below, we will trace the history that has brought us to this point 
where we can meaningfully measure the redshift $z_{\rm max}$ at which 
$d_A(z)$ attains its maximum and use it to test various cosmological 
models. This is a unique aspect of the angular diameter distance, which 
none of the other observable signatures possess. To gauge its impact, 
consider that $z_{\rm max}=\infty$ in a model such as the Milne universe 
(see, e.g., Vishwakarma 2013; Chashchina and Silagadze 2015), so one need 
only show that $z_{\rm max}$ is finite---no matter what its actual value 
is---in order to rule out this cosmology.

Rather than pre-assuming a parametric form for $d_A(z)$ based on 
each individual cosmology, we will employ Gaussian processes (GP; Rasmussen 
and Williams 2006; Holsclaw et al. 2010; Seikel et al. 2012; Yennapureddy 
and Melia 2017, 2018a, 2018b) to reconstruct the angular diameter distance to the
compact quasar cores in a model-independent fashion. This has
the dual benefit of (1) permitting us to compare the measured
and predicted values of $z_{\rm max}$ without any bias, and
(2) to reliably measure $z_{\rm max}$ even if it turns out that none
of the models considered here is actually correct.  

In addition, to avoid the possibility that a model may match the
value of $z_{\rm max}$ rather well, while still not adequately accounting 
for the overall reconstructed angular-diameter distance, we shall also
carry out model selection based on the optimization of $d_A(z)$ over
the observed redshift range. Though related, these two diagnostics 
do not overlap completely. Note, e.g., that while fitting $d_A(z)$
to the reconstructed curve requires an optimization of the Hubble
constant $H_0$, a determination of $z_{\rm max}$ is completely
independent of the measured expansion rate. Thus, a consistent
prioritization of the models using these two approaches will
be more robust than model selection based on either of them
alone. The model testing based on a comparison of the predicted
and reconstructed $d_A(z)$ functions will be facilitated by our
recent development of a differential area statistic introduced
for this purpose in Yennapureddy \& Melia (2018a, 2018b).

In \S~2 of this paper, we will summarize the various stages of
development behind the approach of measuring $d_A(z)$, 
highlighting the critical steps that have produced a sample of sources
whose compact cores may be used as standard rulers. We then
briefly describe in \S~3 the Gaussian processes approach that will allow
us to analyze these data in a cosmology independent way. The data
and our method of analysis are presented in \S~4, and we use the
measured value of $z_{\rm max}$ to compare seven different models
in \S~5. We will discuss the results based on the analysis
of $z_{\rm max}$  in \S~6 and then independently carry out model selection 
based on how well the angular-diameter distance $d_A(z)$ compares with 
the GP reconstruction over the entire redshift range in \S~7. For the
tests we carry out in this paper, we do not need to know the actual
size of the compact structure. It will be interesting to see how
its recent measurement (Cao et al. 2017b) impacts our interpretation of 
the Hubble constant for each model, however, and we discuss this outcome
in \S~8. We end with our conclusions in \S~9.

\section{Background}
An early suggestion to optimize cosmological parameters by
using the angular diameter distance to compact radio sources 
assumed to have a fixed reference length was made by Kellermann
(1993), whose analysis argued for a cosmological density close to
its critical value. This claim, however, appeared to have been
premature given the possible influence of source evolution with
redshift. Krauss and Schramm (1993) demonstrated that the 
position of $z_{\rm max}$ depends sensitively on the parameters,
particularly the density of dark energy. They concluded that even
an evolution of less than $30\%$ in source size at $z\lesssim 2$
could completely alter the outcome, thereby recommending that one
must conclusively rule out source evolution in order to use
a measurement of $z_{\rm max}$ as a reliable tool for
cosmology.

One possible way to mitigate the impact of source evolution is to 
base the standard ruler on the separation of quasar pairs (Phillipps
1994), but to establish significant constraints on the cosmological
parameters one needs a sample of several hundred pairs of
physically related quasars at $z>1$. Even then, one needs
to have an accurate assessment of the distribution of pair
separations. Phillipps et al. (2002) also developed a variation
on this theme, considering a proximity effect based on the
observed deficit of Ly$\alpha$ forest lines at redshifts close
to that of the illuminating quasar.  The size of the deficient
region is presumably related to the central engine's absolute
luminosity. Unfortunately, this effect is difficult to disentangle
from other factors, such as the intergalactic ionizing flux. 

A concerted effort to better understand the evolution of radio
sources and its impact on the measurement of an angular diameter
distance was therefore initiated by Gurvits (1994) and Kayser (1995),
who attempted to estimate how much evolution was actually
occurring as a function of redshift and to what degree this affected
the optimization of the model parameters. Gurvits modeled the
luminosity dependence of the compact source size as
$l\sim L^\beta(1+z)^n$, and obtained a best fit with $\beta=0.26\pm
0.03$ and $n=-0.30\pm0.90$. As we shall see shortly, however, 
a persistent complication with compact radio sources is that they 
comprise a mixture of quasars, BL Lacs, OVVs, and others, so 
systematic differences among them cannot be so easily disentangled
from actual cosmological variations. Not surprisingly, therefore, this
initial attempt was not very successful in identifying the correct
cosmological model. This point was amplified by Kayser, who
carefully re-analyzed the VLBI compact source data taking into
account biasing of the sample from the limited resolution, and
concluded at that time that a measurement of the angular
diameter distance to these objects could not be used to differentiate
between different models. Dabrowski et al. (1995) also pointed out
that relativistic beaming cannot be ignored in such sources, since
a flux-limited sample of them contains a projected-size distribution
that is biased. Without a more careful identification of an appropriate
sample, this approach therefore has a tendency to produce null results.

The unknown mixture of different quasar types and their possible
evolution with redshift also affected the precision with which Jackson
and Dodgson (1997) could use the angular diameter distance to
compact cores to identify the importance of dark matter in the expansion
dynamics. Nonetheless, their analysis of the angular diameter size versus
redshift relation for 256 ultracompact sources with $0.5<z<3.8$ did
reveal a preference for a cosmology containing dark energy over one
based solely on cold dark matter.

The first evidence that $d_A(z)$ might not increase indefinitely emerged 
in a study of double-lobed quasars within the redshift range $1.0\lesssim z
\lesssim 2.7$ (Buchalter et al. 1998). The apparent angular size of these
objects remained more or less constant with angular-diameter distance,
as one might crudely expect in most Friedmann-Robertson-Walker (FRW) cosmologies
without any significant evolution (see fig.~1 below). The well-known
exception is the Milne universe, for which $d_A(z)$ increases with redshift
everywhere. Similar constraints using radio galaxies as standard rulers
were obtained by Podariu et al. (2003). In both cases, however, the
results were weaker than those based on other kinds of observation,
e.g., Type Ia SNe.

Unlike compact cores, quasars with extended jets are
subject to long-term dynamical evolution and the possible influence of
environment on their structure extending over galactic scales.  
A more specific component within the jets, i.e., shocks whose linear
diameter could in principle be a standard ruler (Wiik \& Valtaoja 2001),
was promising, although the need to estimate their extent by monitoring
their flux density and measuring light travel times, constituted a heavy
reliance on  the pre-assumed cosmology. The outcome was therefore
compliant to the model via a somewhat circular argument. Much of
the subsequent work with radio sources since that time has therefore
focused on trying to reduce the scatter in the angular diameter
size versus redshift relation for compact radio sources.  And one of
the earliest breakthroughs in this direction was made by Gurvits, Kellermann
\& Frey (1999), who studied 330 5 Ghz VLBI contour maps (see also Frey 
1999) in the redshift range $0.011<z<4.72$,
demonstrating that the dispersion in this relation could be significantly
reduced by restricting the sample to only those compact regions
with a spectral index $-0.38\le\alpha\le0.18$ and a total luminosity
density $Lh^2\ge 10^{26}$ W Hz$^{-1}$ (with $h$ the Hubble constant
in units of $100$ km s$^{-1}$ Mpc$^{-1}$). This constraint on the
spectral index still appears to be valid today and we shall use it, along
with more recently developed criteria, to arrive at the sample used
in this paper.

Our current view of jet launching in quasars and radio galaxies
suggests that the base emission is dominated by self-absorbed
synchrotron emission (Blandford \& K\"onigl 1979; Melia \&
K\"onigl 1989), creating optically-thick features with angular 
diameters in the milliarcsecond (mas) range. At typical distances,
these cores extend over roughly 10 parsecs. It is therefore likely
that such small features are influenced very little by the large-scale
environment of the parent galaxies and kpc-scale jets,
so their physical attributes should be similar from source to source 
and reasonably stable over the time they are seen (Kellermann 1993; 
Jackson 2004, 2008). It is thought that the morphology and kinematics of 
compact quasar cores are controlled by only a handful of parameters 
associated with the central engine itself, including its mass (and possibly 
the spin). Therefore, since opaque features in compact quasar cores
typically last only tens of years (Gurvits, Kellermann \& Frey 1999), 
they are expected to be free of long-term evolutionary effects in 
the active galactic nucleus (AGN) where they are found.

These theoretical ideas, along with the identification by
Gurvits, Kellermann \& Frey (1999) of a sub-sample of quasar cores with a reduced
scatter in their apparent size, have generated increasing interest
in using the latter as standard rulers. Lima \& Alcaniz (2002) constrained
the cosmic equation of state with this approach, assuming a flat 
FRW model driven by matter and dark energy. Like several other
ensuing efforts, however, their model fits were characterized by
rather large values of $\chi^2_{\rm dof}$ (the $\chi^2$ per
degree of freedom), indicating that the scatter in the sample
was still too large to draw any definitive conclusions. In
retrospect, it is not surprising that their optimized values of the 
model parameters are not a good match to the latest consensus
(i.e., {\it Planck}) measurements. Similar work by Zhu \&
Fujimoto (2002) and Chen \& Ratra (2003), using the sample
of Gurvits, Kellermann \& Frey (1999), attempted to
constrain model parameters in cosmologies with a variable
dark energy component and produced interesting limits,
though weaker than those based on other methods.

An early attempt at using angular diameter distance
measurements of milliarcsecond compact quasar cores
to test alternative cosmological expansion scenarios
was carried out by Jain, Dev \& Alcaniz (2003), who
simply modeled the expansion factor as a power law
in time ($a\sim t^\beta$), and concluded that the
data at that time favoured a cosmic evolution with
$\beta=1$. Without the refinements we will discuss
below, however, their sample had too much scatter
and the errors were simply too large for them to say
anything definitive about the value of $z_{\rm max}$. 
Nonetheless, it is interesting to note that even with
the inferior set of data at their disposal, their analysis
seemed to indicate that a linear expansion was favoured
by the observations. As we shall see later in this paper,
our comparison of various cosmologies using compact
quasar cores will demonstrate that the $R_{\rm h}=ct$
universe is favoured over the other models. This cosmology
features a linear expansion, so our results appear to be
consistent with those of Jain, Dev \& Alcaniz (2003).

The sample of ultracompact radio sources often used
in such studies today is that assembled by Jackson \&
Jannetta (2006), extracted from an old 2.29 GHz
VLBI survey of Preston et al. (1985) and additions by
Gurvits (1994). Their own application of these data to construct
the angular diameter versus redshift diagram produced
results more in line with those based on observations
of the CMB, including a measurement of the angular
size of the acoustic horizon. We ourselves will also use this catalog
as the basis of our analysis, though with several critical
improvements that we shall discuss shortly. These
refinements are necessary because, as alluded to earlier,
a persistent complication with ultracompact cores
is that they constitute a mixed population of 
AGNs---quasars, BL Lacs, OVVs, etc.---making it difficult to 
disentangle systematic differences from true cosmological 
variations. 

The recent introduction of an additional luminosity restriction 
applied to sources in the compact core sample, used in conjunction 
with the constraint on the spectral index $\alpha$ used earlier by 
Gurvits, Kellermann \& Frey (1999), appears to have overcome
this weakness. These authors, and independently Vishwakarma (2001),
had already shown that the exclusion of sources with low luminosities
$L$ could mitigate the dependence of the intrinsic core size on $L$
and redshift $z$. In their analysis of the Jackson \& Jannetta (2006)
sample, Cao et al. (2017a, 2017b) have demonstrated 
a strong dependence of the core size ${\ell}_{\rm core}$ on luminosity, not just 
at the low end (as had been noted earlier), but also at the high end as well
(see also Cao et al. 2015; Li et al. 2016; Zheng et al. 2017). Adopting the
parametrization ${\ell}_{\rm core}={\ell}_0\,L^\gamma(1+z)^n$, where
${\ell}_0$ is simply a scaling constant, they showed that only a sub-sample 
of intermediate-luminosity radio quasars in the range
$10^{27}\,{\rm W/Hz}<L<10^{28}\,{\rm W/Hz}$ have a core size with
negligible dependence on luminosity and redshift. For these sources, 
$\gamma\approx 10^{-4}$ and $|n|\approx 10^{-3}$. Therefore,
it appears that a sub-sample selected from the Jackson \& Jannetta
catalog with a restricted spectral index $\alpha$ and luminosity $L$
constitutes a compilation of compact radio cores with a reliable 
standard linear size.

This is the procedure we shall follow in this paper to measure 
$z_{\rm max}$ with unprecedented accuracy and to compare 
cosmological models in ways not previously possible using other 
measures of cosmological distance. In addition, we will avoid
any possible biasing of the results by reconstructing the angular
diameter distance of the compact cores using Gaussian processes (GP),
which we now describe.

\section{Gaussian Processes}
Most fitting procedures require the pre-assumption of a parametric form for
the fitting function tailored to the specific properties of the cosmological model.
The Gaussian Processes (GP) approach (Seikel et al. 2012) avoids this shortcoming
and is thereby not subject to the possibility that the predicted signature may,
or may not, be a reasonable representation of the actual redshift-dependence
of the measurements.\footnote{The full details of this implementation may be
found in Seikel et al. (2012), and a catalog of useful algorithms is
maintained at http://www.acgc.uct.ac.za/ seikel/GAPP/index.html} 

To model a function f(x) rigorously without relying on  any prior parametric 
form, the GP procedure assumes that the $n$ observations of a data set
y=$\{y_1,y_2,....,y_n\}$ are sampled from a multivariate Gaussian distribution. 
The mean of the GP partnered to the data is taken to be zero. Note, however,
that while modeling the data with GP is straightforward, one must face two
potential areas of ambiguity with this technique. We describe these here and
present steps we have developed to ensure that the outcome of the
reconstruction is not affected significantly by our choice of GP components. 

The first of these arises because the values of the function evaluated at 
different points $x_1$ and $x_2$ are not independent of each other. One
must therefore introduce a covariance function $k(x_1,x_2)$ to deal
with the linkage between neighboring points. The difficulty is that
$k(x_1,x_2)$ is not unique or well known. There is often a broad range 
of such covariances. Indeed, while it  makes sense to pick a function
that depends only on the distance between neighboring points, this is 
actually not required. Most applications of this work adopt a squared 
exponential, 
\begin{equation}
k(x_1,x_2) = \sigma_f^2\exp\left(-{(x_1-x_2)^2\over 2l^2}\right)\;,
\end{equation}
which is infinitely differentiable and useful for reconstructing both the
function representing the data and its derivative. 

The so-called hyperparameters $\sigma_f$ and $l$ are not parameters
in the usual sense, since they do not specify the form of the function, 
but rather its `bumpiness.' The length $l$ characterizes the distance in $x$ 
corresponding to a significant variation of the reconstructed function.
The dependence in the ordinate direction is scaled by the signal variance 
$\sigma_f$. Equation~(1) for $\{x_1,x_2,....x_n\}$ observation points 
leads to the covariance matrix
\begin{equation}
K=
\begin{bmatrix}
k(x_1,x_1) & k(x_1,x_2) & ..... & k(x_1,x_n)\\
k(x_2,x_1) & k(x_2,x_2) & ..... & k(x_2,x_n)\\
.......... & .......... &...... & ..........\\
k(x_n,x_1) & k(x_n,x_2) & ..... & k(x_n,x_n)
\end{bmatrix}\;.
\end{equation}
The introduction of a new observation point $x_*$ requires the evaluation
of the vector
\begin{equation}
K_*\equiv
\begin{bmatrix}
k(x_*,x_1) & k(x_*,x_2) & ..... & k(x_*,x_n)
\end{bmatrix}\;,
\end{equation}
and the quantity $K_{**}\equiv k(x_*,x_*)$. Given that the data are
assumed to be represented as a sample from a multivariate GP, 
\begin{equation}
\begin{bmatrix}
y \\ y_* 
\end{bmatrix}
= N\bigg(0,
\begin{bmatrix}
 K & K_*^T \\
 K_*   & K_{**}
\end{bmatrix}
\bigg)\;,
\end{equation}
the reconstruction entails the maximization of the conditional probability
\begin{equation}
p(y_*|y)\sim N(K_*K^{-1}y,K_{**}-K_*K^{-1}K_*^T)\;.
\end{equation}
This distribution has a mean $y_*$, which is given as
\begin{equation}
\mu(y_*)=K_*K^{-1}y\;,
\end{equation}
with a corresponding uncertainty
\begin{equation}
 {\rm var}(y_*)=K_{**}-K_*K^{-1}K_*^T\;.
\end{equation}

The reconstructed function shown in figure~1 below is based on the 
use of the kernel in Equation~(1). To ensure that our measurement
of $z_{\rm max}$ is not being unduly affected by this choice of 
covariance function, we also carry out a parallel set of simulations using 
a very different kind of kernel known as a Mat\'ern covariance function, specifically
the one called Mat\'ern92 (Seikel et al. 2012), whose explicit form is
\begin{eqnarray}
k(x_1,x_2) &=& \sigma_f^2\exp\left(-{3|x_1-x_2|\over l}\right)\left(
1+{3|x_1-x_2|\over l}+\right.\nonumber \\
&\null& \left. {27|x_1-x_2|^2\over 7l^2}+
{18|x_1-x_2|^3\over 7l^3}+{27|x_1-x_2|^4\over 35l^4}\right)\;.
\end{eqnarray}
In so doing, we confirm the results of previous workers, which show that
the choice of kernel may change the $p$-values by a few points, though
the outcome of model comparisons is not altered qualitatively. The rank 
ordering of models listed in Table~1 below is completely unaffected by 
the choice of covariance function. 

The second potential ambiguity is associated with the hyperparameters 
themselves. An often used approach is to train them by maximizing the 
likelihood that the reconstructed function reproduces the measured values 
at the data points $x_i$. The caveat is that for a purely Bayesian analysis, 
the hyperparameters should be marginalized over instead of being optimized, 
but for our application (as is commonly the case), the marginal likelihood 
is sharply peaked, so optimization is a good approximation to marginalization. 
The bottom line is that for a set of data such as we have here (see
fig.~1), there is actually no freedom to choose $\sigma_f$ and $l$
separately once we carry through with the optimization procedure
described above.

\section{Data and Analysis}
As we have seen from the above discussion, we may now choose
from the many hundreds of available VLBI images a reduced
sample of quasar cores with manageable scatter and little,
if any, evolutionary effects by excluding those sources with low 
and high luminosities, $L$, and extreme spectral 
indices, $\alpha$. Specifically, the dispersion
in (linear) core size ${\ell}_{\rm core}$ is significantly
mitigated by selecting only sources with $-0.38<\alpha<0.18$ 
(Gurvits, Kellermann \& Frey 1999; Cao et al. 2017a, 2017b), and
an intermediate-luminosity $10^{27}\,{\rm W/Hz}<L<10^{28}\,
{\rm W/Hz}$ (Cao et al. 2017). These two criteria result in a compact 
quasar-core catalog with robust standard linear sizes.

\begin{figure}
\centerline{
\includegraphics[angle=0,scale=0.72]{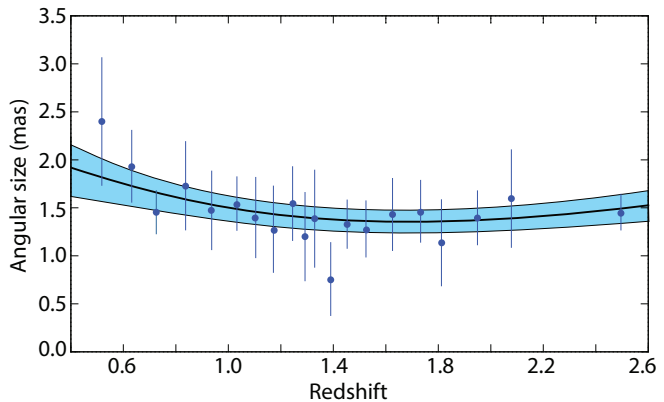}}
\vskip 0.02in
\caption{Angular size of 140 compact quasar cores divided into
bins of 7, as a function of redshift. Each datum represents
the median value in its bin. The thick solid curve is the 
reconstructed angular-size function using Gaussian processes, 
while the shaded region shows the $1\sigma$ variation.}
\end{figure}

The data we use here are drawn from the 613 sources assembled 
by Jackson \& Jannetta (2006) using the 2.29 GHz VLBI survey 
of Preston et al. (1985) and additions by Gurvits (1994). 
We use the {\it Planck} optimized parameters (Planck Collaboration 
2016) to estimate the luminosity distance which, together
with the measured total flux density at $2.29$ GHz, gives the
luminosity $L$, which may be used to extract the subsample 
with intermediate luminosities. In doing so, we are giving 
{\it Planck} $\Lambda$CDM the benefit of the doubt, but note 
that this procedure is used merely to estimate $L$; these 
parameters are not used in any
other way during the model comparisons described below,
so the results are not biased by this approach. Reducing
the sample further by restricting the range of $\alpha$ 
produces a final catalog of 140 sources for our analysis. 
We bin these sources into groups of 7 and select the median
value in each bin to represent the angular size (Santos 
\& Lima 2008). We take this step to partially minimize an
additional degree of scatter that would otherwise appear
using the individual data points. The resulting $20$ data 
points are plotted in figure~1, along with their $1\sigma$ 
errors estimated assuming Gaussian variation within each bin.
The caveat with this approach is that the scatter may not
be purely Gaussian, e.g., if there is some contribution from
an unspecified systematic effect. In future work, we will
address this question using a two-point diagnostic method
we have already applied to other kinds of data, such as
$H(z)$ versus $z$ (Leaf \& Melia 2017a) and the HII galaxy 
Hubble diagram (Leaf \& Melia 2017b). As we have shown in 
these previous applications, the two-point diagnostics very 
effectively indicate the quality of the errors and their likely 
contributions.

The angular-diameter distance (solid curve) reconstructed with 
Gaussian processes (GP; Rasmussen \& Williams 2006; Holsclaw
et al. 2010; Seikel et al. 2012; Yennapureddy \& Melia 2017, 2018a, 2018b) 
allows us to study the geometry 
of the Universe in a new, unique way. Compact radio cores have been 
mapped with VLBI as far out as $z\sim 4$, allowing us to probe 
the geometry of the Universe over $80\%$ of its existence. This 
happens to be the redshift range within which $d_A(z)$
first increases, reaches a maximum at some $z_{\rm max}$, and then 
shrinks to zero as $z\rightarrow \infty$. The physics behind this
phenomenon is actually easy to understand (Melia 2013). The 
angular-diameter distance is based on the measurement of a lateral 
proper size, so we see the object in projection as it was when it 
emitted the light approaching us today. But all sources were closer 
to us as we look back in cosmic time, so the {\it apparent} angular size 
$\theta_{\rm core}$ of compact quasar cores actually increases as 
$z\rightarrow\infty$, meaning that $d_A(z)$ ($\sim \theta_{\rm core}^{-1}$) 
therefore gets smaller.

\begin{figure}
\centerline{
\includegraphics[angle=0,scale=0.45]{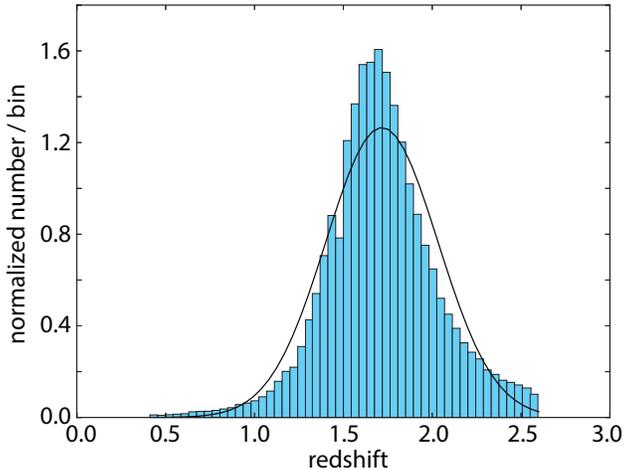}}
\vskip 0.02in
\caption{Distribution of $z_{\rm max}$ values calculated from 50,000
mock samples generated from the data and errors shown in fig.~1
(see text). The histogram is truncated at $z\sim 2.6$ due to a 
lack of data beyond this point. Modeling this distribution as an
approximate Gaussian, we find that $68.2\%$ of the realizations lie
within a standard deviation $\sigma_{z_{\rm max}}=0.20$ of the measured 
value $z_{\rm max}=1.70$.}
\end{figure}

We emphasize that as long as $\ell_{\rm core}$ is a true standard ruler, at least 
in an average sense, we do not need to know its actual value to identify
$z_{\rm max}$ because we are simply sampling the ratio of scales at 
different redshifts. We also do not need to know the Hubble constant, $H_0$, 
which does not affect the location of the turning point in $d_A(z)$. The
net result is that the GP reconstruction shown in figure~1 is completely
free of any cosmological model and assumptions. The turning point
$z_{\rm max}$ may then be used unambiguously to test the predictions 
of the models described below.

To find the error associated with this measurement of $z_{\rm max}$,
we adopt the following procedure. We use the data and their $1\sigma$ 
errors shown in figure~1 to create mock samples of $20$ values of the 
core size, $\theta_i\equiv\theta(z_i)$, with the same redshifts, $z_i$ 
($i=1,...,20$), as the actual measurements, but with Gaussian randomized 
values $\theta_{\rm mock}(z_i)=\theta(z_i)+r\sigma_i$, where $r$ is a
Gaussian random variable with mean $0$, and variance $1$, and $\sigma_i$
is the dispersion at $z_i$.

Then, for each mock sample, we redo the Gaussian-process reconstruction 
to find its corresponding $z_{\rm max}$, and repeat this process 50,000
times. The distribution of $z_{\rm max}$ values thus constructed is shown 
as a histogram in figure~2. This distribution approximates a Gaussian,
but not completely given that the data are truncated at $z\sim 2.6$.
Its mean redshift is consistent with our measured value $z_{\rm max}= 1.70$ 
at the maximum $d_A$ (i.e., the minimum $\theta_{\rm core}$). Crucially, 
approximating this distribution as a Gaussian (shown as a solid black curve
in fig.~2), we find that $68.2\%$ of the realizations occur within a 
standard deviation $\sigma_{z_{\rm max}}=0.20$, which we adopt as a 
reasonable estimate of the measurement error for $z_{\rm max}$.

\section{Cosmological Models}
We will compare the measured value of $z_{\rm max}$ with the prediction of
seven different cosmological models, each with its unique expression for the 
angular-diameter distance, $d_A(z)$. As noted, the Hubble constant does not 
affect $z_{\rm max}$, so we do not need to specify its value. 

\begin{enumerate}

\item The flat {\it Planck} $\Lambda$CDM model, with parameters 
$\Omega_{\rm m}$, $\Omega_\Lambda$ and a dark-energy equation-of-state 
$w_\Lambda =-1$. In the following expressions, $\Omega_i$ is the 
energy density of species $i$, scaled to today's critical density, 
$\rho_c\equiv 3c^2H_0^2/8\pi G$. For this model,
\begin{equation}
d_A(z)=\frac{c}{H_0}{1\over1+z}
\int _0^z\frac{du}{\sqrt{\Omega _m(1+u)^3
+\Omega _\Lambda (1+u)^{3(1+w_\Lambda )}}}\;.\label{concordance}
\end{equation}
For the {\it Planck} parameters (Planck Collaboration 2016)
$\Omega_{\rm m}=0.308\pm 0.012$ and $\Omega_\Lambda=1.0-\Omega_{\rm m}$,
we find that $z_{\rm max}=1.594$ for this model.

\item Einstein--de Sitter (i.e., Eq.~1 with $\Omega_{\rm m}=1$ and
$\Omega_\Lambda=0$):
\begin{equation}
d_A(z)=2\frac{c}{H_0}{1\over 1+z}\left(1-\frac{1}{\sqrt{1+z}}\right)\;.
\end{equation}
As with several other models introduced below, this cosmology is
disfavoured by many other observations (though see Vauclair et al.
2003; Blanchard 2006), but we include it in this list because the 
measurement of $z_{\rm max}$ provides an important complementary
(and unique) measure of the Universe's geometry. The angular-diameter distance
in this model attains its maximum value at $z_{\rm max}=0.682$.
\vskip 0.1in

\item The $R_{\rm h}=ct$ Universe (a Friedmann-Robertson-Walker cosmology
with zero active mass; Melia 2016a, 2017a). In this model, the total equation-of-state
is $\rho+3p=0$, in terms of the energy density $\rho$ and pressure $p$ 
(Melia 2007; Melia and Shevchuk 2012). In this case, 
\begin{equation}
d_A(z)=\frac{c}{H_0}{1\over 1+z}\ln (1+z)\;,
\end{equation}
and $z_{\rm max}=1.718$.
\vskip 0.1in

\begin{table*}
\center
{\footnotesize
\centerline{{\bf Table 1.} $z_{\rm max}$ for seven cosmological models}
\begin{tabular}{lcrr}
\\
\hline\hline
\\
{Model}\qquad& $z_{\rm max}$ & $|\,z_{\rm max}-z_{\rm max}^{\rm obs}\,|\,/\,\sigma$
& Probability ($\%$)\\
\\
\hline
\\
$R_{\rm h}=ct$ & $1.718$& $0.09$\qquad\quad\null & $92.8$\qquad\null \\
{\it Planck} $\Lambda$CDM & $1.594$ & $0.53$\qquad\quad\null & $59.6$\qquad\null \\
Einstein-de Sitter & $0.682$ & $5.09$\qquad\quad\null & $\sim 0$\qquad\null \\ 
Milne universe & $\infty$ & $\infty$\qquad\quad\null & $0$\qquad\null \\
Static Euclidean & $\infty$ & $\infty$\qquad\quad\null & $0$\qquad\null \\
Static Euclidean tired light & $\infty$ & $\infty$\qquad\quad\null & $0$\qquad\null \\
Static Euclidean plasma tired light\qquad\qquad & $\infty$ & $\infty$\qquad\quad\null & 
        $0$\qquad\null \\
\\
\hline\hline
\end{tabular}
}
\vskip 0.3in
\end{table*}

\item The Milne Universe. This (well-known) solution is also an
FRW cosmology, though with an energy density, pressure and cosmological
constant all equal to zero. Its spatial curvature is negative ($k=-1$).
It follows from the Friedmann equations that the scale factor is
linear in time (see, e.g., Vishwakarma 2013; Chashchina and Silagadze 2015).
Note, however, that although it shares this linear expansion with the
$R_{\rm h}=ct$ universe, the observable signatures in these two
models are different because---unlike Milne---the latter is not an
empty universe. In Milne, the angular-diameter distance is
\begin{equation}
d_A(z)=\frac{c}{H_0}{1\over 1+z}\sinh\left[{\rm ln} (1+z)\right]\;,
\end{equation}
which should be contrasted with Equation~(11) for $R_{\rm h}=ct$. Note
that this expression for the angular-diameter distance has no turning
point, so for Milne $z_{\rm max}=\infty$.
\vskip 0.1in

\item Static Euclidean cosmology with a linear Hubble law at all redshifts:
\begin{equation}
d_A(z)=\frac{c}{H_0}z\label{angdistst}\;.
\end{equation}
This model, which assumes that the Universe is static, has been applied to 
certain specific observations (Lerner et al. 2014). The factor $\sqrt{1+z}$ 
arises from the loss of energy due to a redshift without expansion. 
It differs from the more commonly found factor $(1+z)$ because in this
model there is no time dilation. Of course, there are significant 
challenges in finding consistency between this scenario and other
kinds of data, but in this paper, our goal is simply to test its
predicted value of $z_{\rm max}=\infty$ (like Milne) 
against the measurement. 
\vskip 0.1in

\item Static Euclidean cosmology with tired light:
\begin{equation}
d_A(z)=\frac{c}{H_0}{\rm ln} (1+z)\;. 
\label{angdisttl}
\end{equation}
This phenomenological model assumes that photons lose energy due
to some interaction along their trajectory, and that this
loss of energy scales as the path length, i.e., 
${dE}/{dr}=-(H_0/c)E$ (LaViolette 2012). Of course, as in the previous 
model, this ansatz is not very successful in accounting for
many other observations, but our goal here again is to simply focus 
on the unique observational signature $z_{\rm max}=\infty$
(again, like Milne).
\vskip 0.1in

\item Static Euclidean model with plasma tired light:
\begin{equation}
d_A(z)=\frac{c}{H_0}\ln (1+z)\;.
\end{equation}
In this plasma redshift application, there is an additional 
Compton scattering that is double that of the plasma redshift 
absorption (Brynjolfsson 2004; \S 5.8). As in Milne and the
previous tired light models, we find that $z_{\rm max}=\infty$.
\end{enumerate}

\section{Results based on $z_{\rm max}$}
Table~1 compares each model's prediction with the measured $z_{\rm max}$,
along with the difference as a fraction of $\sigma_{z_{\rm max}}$ (the 
so-called $z$-value), and the corresponding probability that the
predicted turning point is consistent with its measured value. Note
that these percentages are absolute; in other words, we are not comparing 
relative probabilities, so they do not necessarily add up to one.
Each individual model's comparison with the data is independent of the 
relative merits of the other cosmologies.

The surprising feature of Table~1 is how strongly the various models
are differentiated on the basis of $z_{\rm max}$ alone, even without
considering other kinds of data. One of the most important aspects
of the expansion dynamics tested by Type Ia SNe is the hypothesized
transition from deceleration to acceleration at $z\sim 0.7$. Now
we see that compact quasar cores play an equally important role 
in examining the geometry of the Universe at another critical
transition redshift, $z_{\rm max}$, where the angular-diameter
distance turns over. This characteristic redshift is so different
between competing cosmologies that its measurement already 
favours only two of the models we examine here, principally
$R_{\rm h}=ct$, followed by {\it Planck} $\Lambda$CDM.

But in spite of {\it Planck} $\Lambda$CDM not being the
preferred model, there is enough flexibility
in the expression for $d_A(z)$ in $\Lambda$CDM (Eq.~1) that we
should consider whether an alternative set of parameter values 
might raise its standing to that of $R_{\rm h}=ct$. Indeed, a 
variation of flat $\Lambda$CDM with $\Omega_{\rm m}=0.23$
predicts a turning point at $z_{\rm max}=1.70$, consistent with
the measured value, and a probability exceeding $92.8\%$. This 
scaled matter density, however, would be in tension at more than
$6.5\sigma$ with the {\it Planck} optimized value. Additional 
flexibility could be introduced by relaxing the constraint that
dark energy is a cosmological constant, so that $w_{\rm de}\not=-1$. 
But with each such modification to the standard model, we recede 
further and further from the concordance {\it Planck} cosmology, 
calling into question whether finding consistency with the measured 
value of $z_{\rm max}$ is worth damaging the optimization of fits 
to other data, including the cosmic microwave background. 

Aside from this head-to-head comparison between $R_{\rm h}=ct$
and $\Lambda$CDM, the results in Table~1 also strongly support 
other observations that have disfavoured---or even rejected---the 
other five models examined here. For example, note how different 
the outcome is for Milne compared to $R_{\rm h}=ct$. Given that 
both of these models predict a linear expansion rate, they are 
still sometimes confused with each other in the literature. On
occasion, Milne is compared to $\Lambda$CDM to `demonstrate'
that $R_{\rm h}=ct$ is disfavoured by the data (see, e.g., 
Melia 2015, and references cited therein). But as is clearly 
demonstrated here, the observational signatures associated with 
Milne are very different from those in $R_{\rm h}=ct$ and,
while the latter is favoured by the data, the former is strongly
ruled out.

Finally, we comment on one of the issues highlighted in \S~3
concerning a possible influence due to the choice of covariance
function $k(x_1,x_2)$ while using Gaussian processes. As we
alluded to earlier, the choice of $k$ in Equation~(1) is not
unique, though this particular function has found widespread
appeal with GP applications (see, e.g., Rasmussen \& Williams
2006; Holsclaw et al. 2010; Seikel et al. 2012; Yennapureddy
\& Melia 2017, 2018a, 2018b). What determines whether or not a particular
function is appropriate is how far the correlation extends to
either side of each datum, and whether the chosen $k$
adequately models this correlation in terms of the hyperparameter
$l$ (and to a lesser degree $\sigma_f$). We have therefore
tested the correlation in our data set by reconstructing the
angular size function in figure~1 using the very different
kernel in Equation~(8). The results are virtually identical
to those shown in Table~1, except that the individual
percentages change by a few points or less. Very importantly,
the rank ordering of these seven models remains exactly the
same as that shown here. We are therefore confident that
the GP method has been applied correctly to these data. 

\begin{figure}
\centerline{
\includegraphics[angle=0,scale=0.72]{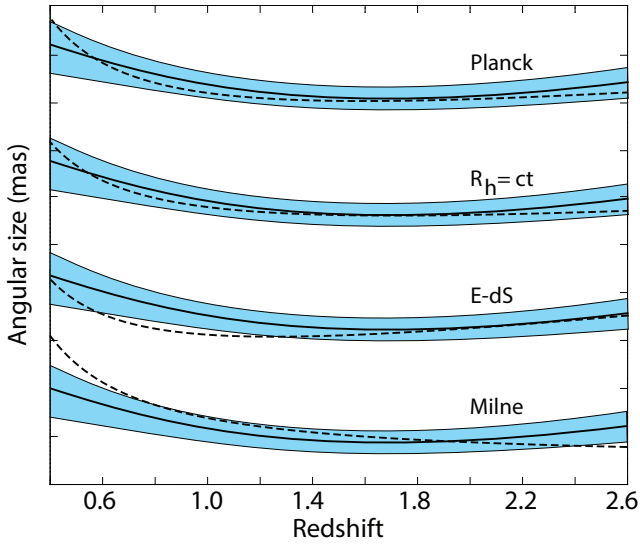}}
\vskip 0.02in
\caption{Comparison of the GP reconstructed quasar core size $\theta_{\rm core}(z)$
(solid curve, from fig.~1) with the model prediction (dashed) for the {\it Planck} 
$\Lambda$CDM, $R_{\rm h}=ct$, Einstein-de Sitter and Milne cosmologies. 
The four plots have been staggered vertically for clarity. In each case, the shaded 
swath represents the $1\sigma$ confidence region for the reconstruction (see fig.~1). 
The cumulative distribution is shown in fig.~5, and the values of $\eta$ (Eq.~17) for 
the optimized fits, along with the corresponding probabilities, are listed in Table~2.}
\end{figure}

\begin{figure}
\centerline{
\includegraphics[angle=0,scale=0.72]{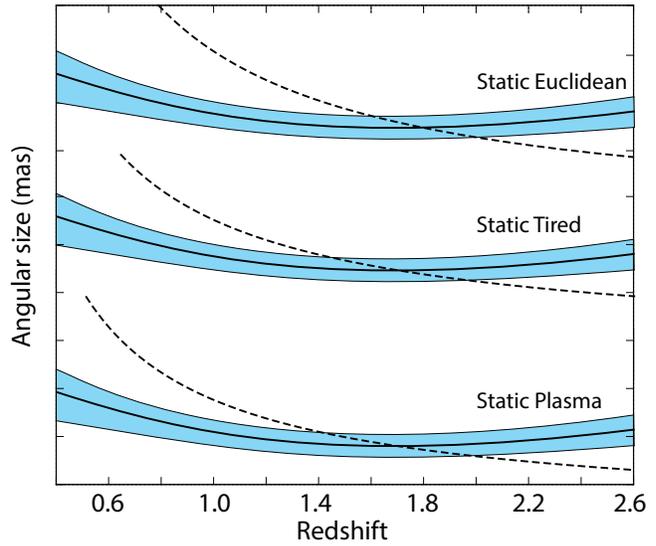}}
\vskip 0.02in
\caption{Same as fig.~3, except now for the Static Euclidean, Static 
Tired Light and Static Plasma Tired Light cosmologies. As indicated in fig.~5 and Table~2, 
these three models are disfavoured by the data in comparison with the 
{\it Planck} and $R_{\rm h}=ct$ models.}
\end{figure}

\section{Results based on $d_A(z)$}
One of the strengths of the $z_{\rm max}$ diagnostic is that
its determination is completely independent of the Hubble constant
$H_0$. Thus all models may be compared with each other on level
ground. But as we alluded to above, there is always the possibility 
that a model's prediction may match the turning point rather well, 
while its angular-diameter distance $d_A(z)$ is overall a poor match 
to the GP reconstructed curve in figure~1. To compare the predicted
and measured $d_A(z)$ functions, however, 
it is necessary to optimize the vertical scaling (proportional to 
$c/H_0$) in equations~(9-15) individually for each cosmology. We
do this by writing the theoretical angular size of the compact quasar 
core as 
\begin{equation}
\theta^{\rm th}_{\rm core}(z) = {\ell_{\rm core}\over d_A(z)}\;,
\end{equation}
where ${\ell}_{\rm core}$ is the physical core size, assumed
to be more or less constant from source to source in the reduced
quasar sample. To carry out the analysis in this paper,
we do not need to know the actual value of $\ell_{\rm core}$ and
$H_0$, and we may combine them by merging the expression for 
$\theta_{\rm core}(z)$ and $d_A(z)$ for each model, writing
\begin{equation}
\theta^{\rm th}_{\rm core}(z) = {\eta\over\mathcal{I}(z)}\;,
\end{equation}
where $\eta\equiv \ell_{\rm core}H_0/c$ and
\begin{equation}
\mathcal{I}(z)\equiv \left({H_0\over c}\right)d_A(z)\;.
\end{equation}
As we shall see shortly, however, a recent measurement
of $\ell_{\rm core}$ by Cao et al. (2017b) allows us to see what the
optimized value of $\eta$ implies for the Hubble constant in each cosmology.
In order to cast each model in its best possible light, we
optimize the parameter $\eta$ individually in each case
to yield the best match with the GP reconstructed curve, 
following a procedure described in Yennapureddy \& Melia (2018a, 2018b),
and briefly summarized below. 

The best fit curves for the models we test in this paper are shown
in comparison with the GP reconstruction and its confidence region (shaded swath)
in figs.~3 and 4. Even a quick inspection by eye shows that Einstein-de Sitter,
Milne, and especially Static Euclidean, Static Tired Light and 
Static Plasma, extend beyond the $1\sigma$ region and are therefore not well 
matched to the data. Planck and $R_{\rm h}=ct$ do much better, 
as reflected also in the probabilities displayed in Table~2.

\begin{figure}
\centerline{
\includegraphics[angle=0,scale=0.72]{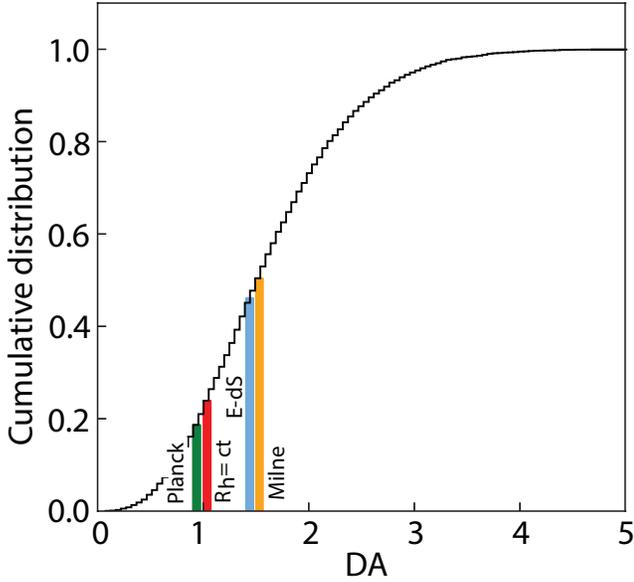}}
\vskip 0.02in
\caption{Cumulative probability distribution for the area differential
$DA$ (Eqn.~20), and the estimated values for the cosmological models
considered in this paper. The two cosmologies favoured by this test
are {\it Planck} and $R_{\rm h}=ct$, all with a probability
$\sim 0.8$, as indicated in Table~2. The other 3 (static) models are 
off the chart to the right, consistent with zero probability (see Table~2).}
\end{figure}

When comparing two continuous functions, i.e., the reconstructed
$\theta_{\rm core}(z)$ curve in figure~1 with $\theta^{\,\rm th}_{\rm core}(z)$
in Equation~(17), one may not use discrete sampling statistics because sampling 
at random points to obtain the differences between the two curves would lose 
information, whose importance is difficult to ascertain. We have recently 
introduced a new diagnostic, called the ``Area Minimization Statistic," to 
estimate the probability that a model is consistent with the data. Assuming 
that the measurement errors are Gaussian, one generates a mock sample of GP 
reconstructed curves covering the likely variation of $\theta_{\rm core}(z)$ 
away from the function representing the actual data. This is done by using 
the Gaussian randomization
\begin{equation}
\theta^{{\rm mock},i}_{\rm core}(z)=\theta^i_{\rm core}(z)+r\sigma_{\theta^i_{\rm core}}\;,
\end{equation}
where $\theta^i_{\rm core}(z)$ are the actual measurements and
$\sigma_{\theta^i_{\rm core}}$ are the corresponding errors. Also, $r$ is a Gaussian
random variable with zero mean and a variance of $1$. GP are then used with these 
$\theta^{{\rm mock},i}_{\rm core}(z)$ mock data and their errors
$\sigma_{\theta^i_{\rm core}}$ to reconstruct the function $\theta^{\rm mock}_{\rm core}(z)$ 
representing each mock sample. In the final step, we determine the weighted
absolute area difference
\begin{equation}
DA= \int_{z_{\rm min}}^{z_{\rm max}}dz\,\frac{\big|\theta^{\rm mock}_{\rm core}(z)-
                \theta_{\rm core}(z)\big|}{\sigma_{\rm GP}(z)}\;.
\end{equation}
Here, $z_{\rm min}$ and $z_{\rm max}$ are the minimum and maximum
redshifts, respectively, of the data range, and $\sigma_{\rm GP}(z)$ is the calculated
dispersion (corresponding to the shaded region in fig.~1) associated with the GP 
reconstructed curve $\theta_{\rm core}(z)$. 
Repeating this procedure $10,000$ times, we build the probability distribution for the
area differential $DA$, which is shown in fig.~5, along with the individually measured values
of $DA$ for each model we test. With the assumption that a smaller $DA$ corresponds
to a better match to $\theta_{\rm core}(z)$, the cumulative distribution may then
be used to estimate the likelihood that the difference between a model's prediction and 
$\theta_{\rm core}(z)$ is due principally to Gaussian randomness, rather than the model
being wrong. A comparison of the likelihoods then prioritizes the
models according to their probability of matching the data. Many statistical
approaches utilize this basic concept, but unfortunately none of the existing 
methods may be used for the comparison of two continuous curves, as we have here.

\begin{table*}
\center
{\footnotesize
\centerline{{\bf Table 2.} DA statistics for seven cosmological models}
\begin{tabular}{lcr}
\\
\hline\hline
\\
{Model}\qquad& $\eta$ & Probability ($\%$)\\
\\
\hline
\\
{\it Planck} $\Lambda$CDM & $0.53$ & $81$\qquad\null \\
$R_{\rm h}=ct$ & $0.48$& $76$\qquad\null \\
Einstein-de Sitter & $0.19$ & $52$\qquad\null \\  
Milne universe & $0.57$ & $50$\qquad\null \\
Static Euclidean & $2.23$ & $0$\qquad\null \\
Static Euclidean tired light & $1.31$ & $0$\qquad\null \\
Static Euclidean plasma tired light\qquad\qquad & $1.31$ 
     & $0$\qquad\null \\
\\
\hline\hline
\end{tabular}
}
\vskip 0.3in
\end{table*}

The model selection based on the angular-diameter distance supports the 
outcome of the $z_{\rm max}$ diagnostic. The two models favoured
by a comparison of the reconstructed and predicted $d_A(z)$ functions,
i.e., {\it Planck} $\Lambda$CDM and $R_{\rm h}=ct$,
are also those most strongly selected according to how well they account
for the observed turning point in $d_A(z)$. But while {\it Planck} 
$\Lambda$CDM and $R_{\rm h}=ct$ are virtually indistinguishable from each
other in fig.~5 and Table~2, their prioritization based on $z_{\rm max}$ is
much stronger (see Table~1). This is largely due to the fact that, while
$\ell_{\rm core}$ and $H_0$ are not used in the identification of the
turning point $z_{\rm max}$, one needs to optimize the ratio of these two 
unknowns in order to find the best fit for $d_A(z)$. This freedom to
optimize $\eta$ makes it easier for a model to fit the angular-size
data, so the differences between the two favoured models are somewhat
mitigated. This is the principal reason we highlighted the $z_{\rm max}$
diagnostic as being superior to $d_A(z)$ for model selection, given that 
it requires no optimization of free parameters, allowing all of the models 
to be compared on level ground.

\section{Discussion}
We have emphasized throughout this paper that knowing 
the actual value of  ${\ell}_{\rm core}$ and $H_0$ is not necessary
to conduct the model comparisons based on the quasar compact core 
data. Attempts at measuring ${\ell}_{\rm core}$ have already met with
some success (Cao et al. 2017b), however, so it would be interesting to see 
what impact this result (${\ell}_{\rm core}=11.03\pm0.25$ pc) has on
the implied value of the Hubble constant $H_0$ via the optimized 
$\eta$ in Table 2.  From the definition of $\eta$ (near Eq.~18), we
infer that $H_0=63.4\pm 1.2$ km s$^{-1}$ Mpc$^{-1}$ in the case
of $R_{\rm h}=ct$, and $H_0=69.9\pm 1.5$ km s$^{-1}$ Mpc$^{-1}$
for $\Lambda$CDM. Both of these are completely consistent with
previously measured values of the Hubble constant in each cosmology.
The confirmation is provided by the latest Planck release (Planck
Collaboration 2016), and the four previous measurements reported 
for $R_{\rm h}=ct$: $63.2\pm1.6$  km s$^{-1}$ Mpc$^{-1}$
(Melia \& Maier 2013); $63.3\pm 7.7$  km s$^{-1}$ Mpc$^{-1}$
(Melia \& McClintock 2015); $62.3^{+1.5}_{-1.4}$  km s$^{-1}$ Mpc$^{-1}$
(Wei, Melia \& Wu 2017); and $63.0\pm 1.2$  km s$^{-1}$ Mpc$^{-1}$
(Melia \& Yennapureddy 2018).

Aside from actually obtaining a value for ${\ell}_{\rm core}$, Cao et al. (2017b)
showed through their analysis, based on measurements of $H(z)$ using cosmic chronometers,
that this length scale is independent of redshift, making it a true standard ruler. 
As long as these measurements are fully model-independent and free of any systematic effects,
the use of compact structure in quasar cores with this ${\ell}_{\rm core}$ may 
constitute a powerful diagnostic for cosmological measurements, such as $H_0$.
Already, we have found complete consistency between this approach and
previous measurements, including the {\it Planck} optimization of model
parameters. 

It should also be remarked that in both cases (i.e., $R_{\rm h}=ct$
and $\Lambda$CDM), the inferred value of $H_0$ disagrees with the
local measurement based on Cepheid variables (Riess et al. 2018).
 It is not yet clear why this happens, but  some authors have speculated
on the possibility that a local `Hubble Bubble' (Shi 1997; Keenan et al. 
2013; Romano 2017; Wei et al. 2017) may be influencing the dynamics within a distance
$\sim 300$ Mpc (i.e., $z<0.07$). If true, such a fluctuation might lead to 
anomalous velocities within this region, causing the nearby expansion to deviate 
somewhat from a pure Hubble flow. Until this issue is resolved, we must rely
on the large-scale measurement of $H_0$ individually for each model. 

\section{Conclusion}
Our ability to measure $z_{\rm max}$ has created an entirely new
probe of the Universe's geometry. In this paper, we
have shown that the predicted value of this turning point changes
considerably between different models, allowing existing measurements,
e.g., of compact quasar-core sizes, to disfavour all but two of 
the models we examined. The model preferred by these
data is $R_{\rm h}=ct$, which also happens to be the cosmology
with the fewest parameters. Indeed, for the purpose of comparing
values of $z_{\rm max}$, this cosmology has no parameters at all,
which makes the consistency between its prediction 
($z_{\rm max}^{R_{\rm h}=ct}=1.718$) and the measured value 
($z_{\rm max}=1.70\pm 0.20$) quite compelling. {\it Planck} 
$\Lambda$CDM is not yet ruled out by these observations, but
in order to bring its prediction in line with the observations, 
one must adopt a scaled matter density $\Omega_{\rm m}=0.23$ 
in tension with {\it Planck} at over $6.5\sigma$.

We have highlighted the measurement of $z_{\rm max}$
as the most probative diagnostic for model selection based on angular
sizes, principally because it relies on fewer parameters than the
angular-diameter distance itself. But to ensure that the predicted
$d_A(z)$ is consistent with the GP reconstructed function for
the preferred models, we have also compared the overall angular-diameter
distance to the data. The two models most highly favoured 
by this comparison, {\it Planck} $\Lambda$CDM and
$R_{\rm h}=ct$, are also those most strongly preferred by 
$z_{\rm max}$. But unlike the latter, a comparison of $d_A(z)$
with the data is less discerning for these two models, 
mainly due to the additional free parameter, $H_0$, which provides 
more flexibility with the best fit. For example, the optimized value 
of $H_0$ is $63.4\pm1.2$ km s$^{-1}$ Mpc$^{-1}$ for $R_{\rm h}=ct$,
and $69.9\pm1.5$  km s$^{-1}$ Mpc$^{-1}$ for $\Lambda$CDM.
Both are fully consistent with previously measured values of the
Hubble constant in each model, though they differ from each
other at about $4\sigma$. In contrast, model selection based
on $z_{\rm max}$ is independent of $H_0$.

All of the comparative tests completed thus far (see, e.g., 
Table~1 in Melia 2017b for a summary and references) suggest that
the zero active mass condition in general relativity is the influence 
guiding the Universe's expansion. Several exciting new tests are
under development, including the measurement of redshift drift
(Melia 2016b) and the detection of progenitors to high-$z$
quasars (Fatuzzo \& Melia 2017). Within a few years, we should
know for certain whether or not $R_{\rm h}=ct$ is the correct
cosmology. The consequences are far reaching. Chief among them
is the fact that, while inflation is necessary to maintain the
internal self-consistency of $\Lambda$CDM, it is not required for
(and is actually inconsistent with) the zero active mass condition
(Melia 2014). 

\section*{Acknowledgments} We are grateful to the anonymous
referee and Louis Marmet for providing a helpful set of comments and 
suggestions that have led to a significant improvement in this manuscript. 
FM is also grateful to Amherst College for its support through 
a John Woodruff Simpson Lectureship, and to Purple Mountain Observatory in Nanjing, 
China, for its hospitality while part of this work was being carried out. This work 
was partially supported by grant 2012T1J0011 from The Chinese Academy of Sciences 
Visiting Professorships for Senior International Scientists, and grant GDJ20120491013 
from the Chinese State Administration of Foreign Experts Affairs.

\label{lastpage}

\end{document}